\documentclass[twocolumn,twoside,slac_two]{revtex4}
\usepackage{graphicx}
\usepackage{fancyhdr}
\usepackage{hyperref}
\hypersetup{
    colorlinks=true,
    urlcolor=magenta,
}
\begin{document}

\title{The JEM--EUSO program}

%

\author{F. Fenu, for the JEM--EUSO collaboration}
\affiliation{INFN Torino, via Pietro Giuria 1, 10125 Torino, Italy}

\begin{abstract}
JEM--EUSO is an international collaboration for the development of space--based Ultra High Energy Cosmic Ray (UHECR) detectors. 
The instrument consists of a wide Field Of View (FOV) camera  for the detection of the UV light emitted by Extensive Air Showers (EAS) in the atmosphere. 
Within the JEM--EUSO framework several pathfinders have been developed or are in course of development: EUSO--TA, EUSO--Balloon, EUSO--SPB and Mini--EUSO. 
For the near future the K--EUSO detector is foreseen to detect cosmic rays from space.
In this paper we present the JEM--EUSO project and give an overview of the pathfinders and of their results.
\end{abstract}

\maketitle

\thispagestyle{fancy}

\section{Introduction}
\begin{figure}
\includegraphics[width=45mm]{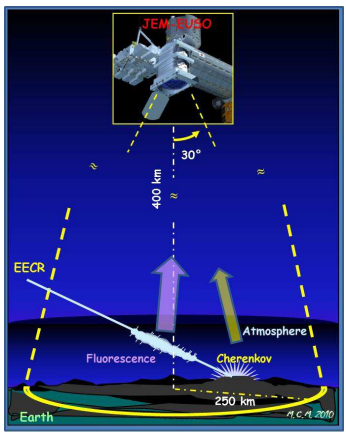}
\caption{the JEM--EUSO observational principle}
\label{fig1}
\end{figure}

The Extreme Universe Space Observatory on board the JEM exposure facility of the ISS (JEM--EUSO) is a detector concept for the study of UHECRs \cite{JEIntro}. 
This mission focuses on the spectrum above 5 $\times 10^{19}$ eV where the flux is extremely low and the required detector areas are extremely large.
The detector consists of a wide FOV ($\pm$ 30 deg) downward--looking UV camera orbiting at 400 km altitude which monitors a $\sim$ $10^{5}$ km$^{2}$ area of atmosphere.
JEM--EUSO aims at the detection of the fluorescence light emitted by EAS in the atmosphere (see Fig. \ref{fig1}). In such a way we can reconstruct the direction, energy and X$_\mathrm{max}$ of the showers at the most extreme energies. 
The identification of the direction of arrival jointly with a proper energy reconstruction is mandatory for anisotropy studies, while the possibility to measure the longitudinal profile of the shower is needed to constrain the average mass at 10$^{20}$ eV.
Another big advantage of this technique with respect to ground based detectors is the uniformity of the sky coverage: JEM--EUSO at 400 km can orbit every 90 minutes the entire earth covering both hemispheres with good uniformity.

The main scientific objectives of the project are: the study of the anisotropies at the extreme energies with unprecedented statistics, the identification of the sources (and possibly the reconstruction of their spectra) and the high statistics measurement of the trans--GZK spectrum \cite{Olinto2015}. 
We define moreover several other exploratory objectives: the search of UHE neutrinos, the search of UHE gamma photons and the study of the galactic and extragalactic magnetic fields through the measurement of the magnetic point spread function of a source.
The nature of the detector allows also to study other phenomena like TLEs, lightnings, the airglow, Auroras, meteors, space debris, bioluminescence in the oceans and to search (or put limits on the flux of) hypothetical phenomena like nuclearites.

JEM--EUSO is a refractor consisting of a system of 3 Fresnel lenses focusing the light on the focal surface \cite{JEInstrument}.
The focal surface is made by $\sim$ 5000 Multi Anode Photomultipliers (MAPMT) produced by Hamamatsu (R11265--M64) capable of single photon counting with a double pulse resolution of the order of $\sim$ 10 ns. The chosen PMT is 2.7 cm side and has 64 pixels each of 3 $\times$ 3 mm.
Four MAPMTs are organized in the so called Elementary Cells (EC) while 2 ECs ie. a block of 6$\times$6 MAPMTs is making a Photo Detection Module (PDM). 
The electronics is modular in order to guarantee high redundancy and organized in several levels: the SPACIROC ASIC for the front end electronics, an FPGA for the PDM--level electronics, the Cluster Control Board and the central CPU.
 The front end electronics counts the photoelectrons within a Gate Time Unit (GTU) window of 2.5 microseconds. The trigger is organized hierarchically and has the challenging task to reduce the trigger rate from $\sim$ 100 GB/s down to the 3 GB/day allowed by the telemetry, still keeping the physical events. 

\section{Prototypes}

The high complexity and risk of this mission demands the development of several pathfinders. We will present here all of them: EUSO--TA which is taking data at the Black Rock Mesa (BRM) site of the Telescope Array (TA) observatory, the EUSO--Balloon (operated by CNES) which flew in August 2014, EUSO--SPB which is foreseen to fly in spring 2017 on a NASA Super Pressure Balloon (SPB) to detect CRs and finally Mini--EUSO, the first space prototype to be launched on the ISS within 2017, whose main goal is to measure the airglow and several atmospheric phenomena visible from space. To conclude, we will present the K--EUSO detector, which is capable of doing UHECR science from space and is foreseen to fly on the ISS by the year 2020.

\subsection{EUSO--TA}
\begin{figure}
\includegraphics[width=65mm]{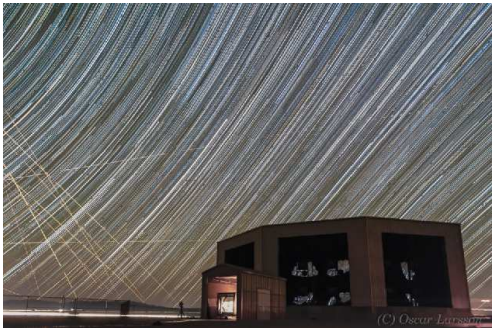}
\caption{the EUSO--TA detector in front of the BRM site of the TA observatory.}
\label{fig3}
\end{figure}

\begin{figure}
\includegraphics[width=65mm]{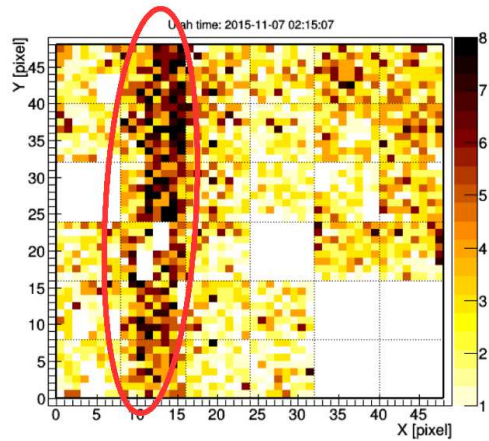}
\caption{ a CR event detected by EUSO--TA thanks to the TA trigger signal. A very preliminary estimate of the energy is 10$^{18.3}$ eV. The event is falling 2.6 km away and EUSO--TA sees with magnified resolution a part of the shower.}
\label{fig4}
\end{figure}

EUSO--TA is a pathfinder of the JEM--EUSO detector \cite{EUSO-TA} which is taking data at the BRM site of the TA observatory in Utah (see Fig. \ref{fig3}). 
This detector consists of two square Fresnel lenses which measure $\sim$ 1 m side and a single PDM of $\sim$ 16 cm side.
The detector is pointing to the sky, overlapping with the TA field of view. 
Each of the 2304 pixels is covering $\sim$ 0.18$\times$0.18 degrees$^2$. The elevation of the telescope can be changed from 0 to 25 degrees. 
The purpose of this pathfinder is the detection of CRs in coincidence with TA. The trigger signal of TA is therefore used as an external trigger and the acquisition of a packet of 128 GTUs is started in coincidence with the TA trigger.
EUSO--TA is detecting in this way several CR events (one example in Fig. \ref{fig4}). Another purpose was the calibration of the detector, also relative to TA.

We used the TA Central Laser Facility (CLF) and other dedicated mobile laser sources to test the technology with events reproducing the kinematics of CR events.
The laser is shot up in the sky and the scattered light was detected by EUSO--TA in such a way as to mimic the signal of a cosmic ray shower.
We also used flashers, flat screens, LEDs to perform a complete characterization of the instrument.
We also detected meteors, lightnings, planes and clouds and therefore we tested the detection and the analysis of such events for JEM--EUSO.
The EUSO--TA detector could be also used to develop the JEM--EUSO autonomous trigger. Indeed EUSO--TA data have been used offline to test and optimize the trigger algorithms. 
As a further step, the PDM board with the autonomous trigger was installed on the EUSO--TA detector and the trigger has been tested online on laser events.
The trigger performances are compliant with the expectations.

\subsection{EUSO--Balloon}
\begin{figure}
\includegraphics[width=85mm]{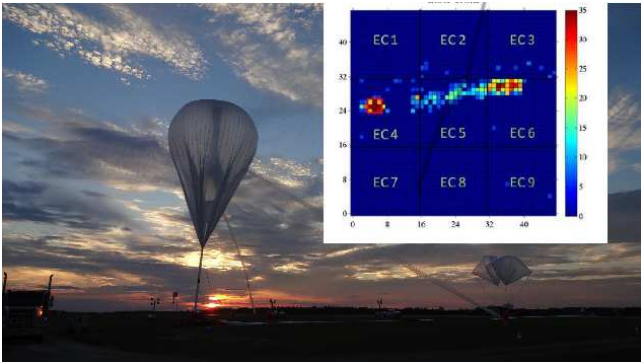}
\caption{EUSO--Balloon launch from Timmins, Canada. August 25$^{th}$ 2014. In the box: the integrated image of a laser shot together with the flasher and LED.}
\label{fig5}
\end{figure}

\begin{figure}
\includegraphics[width=85mm]{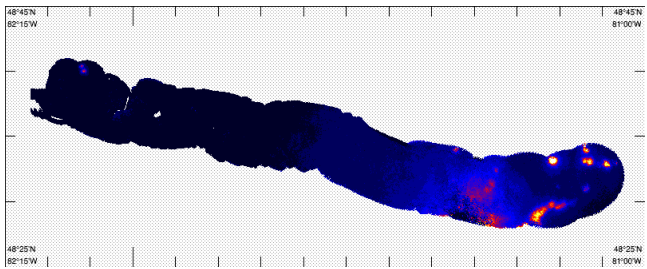}
\caption{the integrated image of the UV background as measured by EUSO--Balloon during the 8 hours flight.}
\label{fig7}
\end{figure}

EUSO--Balloon is a CNES stratospheric balloon pathfinder of the JEM--EUSO mission, which flew on August 25$^\mathrm{th}$ 2014 from Timmins Canada (see Fig. \ref{fig5}). 
The purpose of this pathfinder was to test the JEM--EUSO technology in space environment and to test the response of the detector with respect to several artificial CR events \cite{EUSO-Balloon}. 
The detector consisted of a system of two square, 1 m side Fresnel lenses and of a single PDM.
The flight lasted one night and has successfully proven the capability of the detector to operate in stratospheric conditions (38 km altitude).  
Unexpectedly the detector landed in water also proving the water tightness of the instrument. 
The FOV was 8 $\times$ 8 km$^2$ and each single pixel covered a projected area 120 $\times$ 120 m$^2$ on the ground. 
The balloon covered the distance of roughly 100 km passing over different landscapes including forest lakes, cities, clear and cloudy sky. 
During the flight the balloon was followed at 3,000 m altitude by an helicopter which was equipped with a flasher, a LED and a laser to generate artificial light events (see one example in Fig. \ref{fig5}).
We could therefore detect several hundreds artificial events which were used to test the electronics. 
We used such data to test the algorithms for the reconstruction of the direction of the laser events \cite{eserTrigger}. 
The RMS of the angular reconstruction was estimated in \cite{eserTrigger} to be few degrees.
A map of the background in the UV range has been produced and is shown in Fig. \ref{fig7}. 
Thanks to the present study we could give a preliminary estimate (except the direct airglow component) of the UV background which will be seen by JEM--EUSO on orbit both in clear sky and cloudy conditions.

\subsection{EUSO--SPB}
\begin{figure}
\includegraphics[width=65mm]{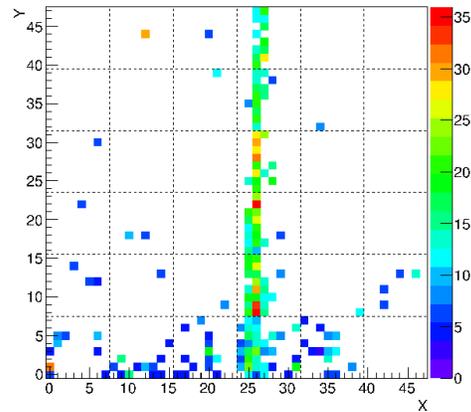}
\caption{the integrated image of a laser shot as seen by EUSO--SPB during the October 2016 testing campaign in Utah. A single pixel threshold has been applied.}
\label{fig9}
\end{figure}

In spring 2017 the EUSO--SPB prototype will be launched from Wanaka, New Zealand \cite{EUSO-SPB}. 
The detector will consist of a system of two Fresnel lenses of 1 m side and one single PDM. The balloon flight will make use of a Super Pressure Balloon (SPB), a new technology currently developed by NASA. 
The test flights performed by NASA in 2015 and 2016 achieved a 32 and 45 days duration.
The detector is assembled and currently (Feb. 2017) in New Zealand in the launch preparatory phase. The detector has now to be fully autonomous and must therefore be provided with solar panels, rechargeable batteries, antennas for data and command transfer.
The main objective of EUSO--SPB is the autonomous detection of CR events for the first time from above using the fluorescence technique. The electronics is therefore equipped with a fully autonomous trigger which has been already tested on the TA site in October 2016. We show in Fig. \ref{fig9} the image of a laser shot which autonomously triggered EUSO--SPB.
The collaboration estimated the expected rate of CR to be of the order of few events during the entire flight.

\subsection{Mini--EUSO}
Mini--EUSO will be launched in 2017 and consists of a system of circular Fresnel lenses and one single PDM to be accommodated inside the ISS \cite{mini-EUSO}.
The detector will be placed in the Russian segment of the ISS behind a UV--transparent downward--looking window and will monitor the atmosphere from 400 km height in the same condition as JEM--EUSO.
The main objective of this prototype is the measurement of the atmospheric UV emission from space.
Given the small diameter of the lens (25 cm), no detection of CRs can be expected. Despite that, the collaboration plans to use artificial sources (with a luminosity equivalent to a 10$^{21}$ eV CR) as in the balloon missions in order to mimic a CR signal.
\begin{figure}
\includegraphics[width=65mm]{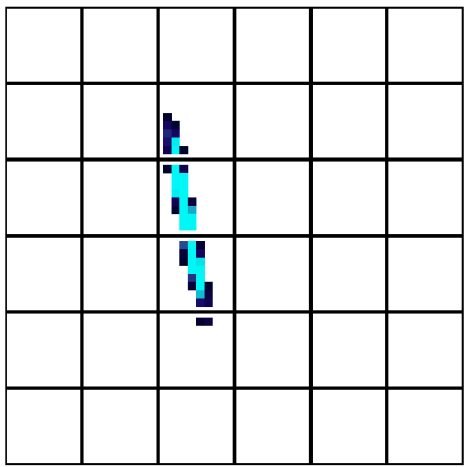}
\caption{the integrated simulated image of a meteor for the Mini--EUSO configuration. A single pixel threshold has been applied.}
\label{fig10}
\end{figure}
Other targets of the missions are the study of TLEs, meteors and auroras. We see in Fig. \ref{fig10} the image of a simulated meteor.
An interesting application of this prototype is the detection of space debris during twilight, which will fly on a lower orbit with respect to the ISS.

\section{K--EUSO}
K--EUSO is the first large size detector being developed in the framework of the JEM--EUSO project \cite{K-EUSO}. The optical design follows a so--called Schmitt optics, namely a combination of a mirror and of a corrector lens. The detector is exploiting the experience gathered in the JEM--EUSO and the KLYPVE collaborations. 
\begin{figure}
\includegraphics[width=85mm]{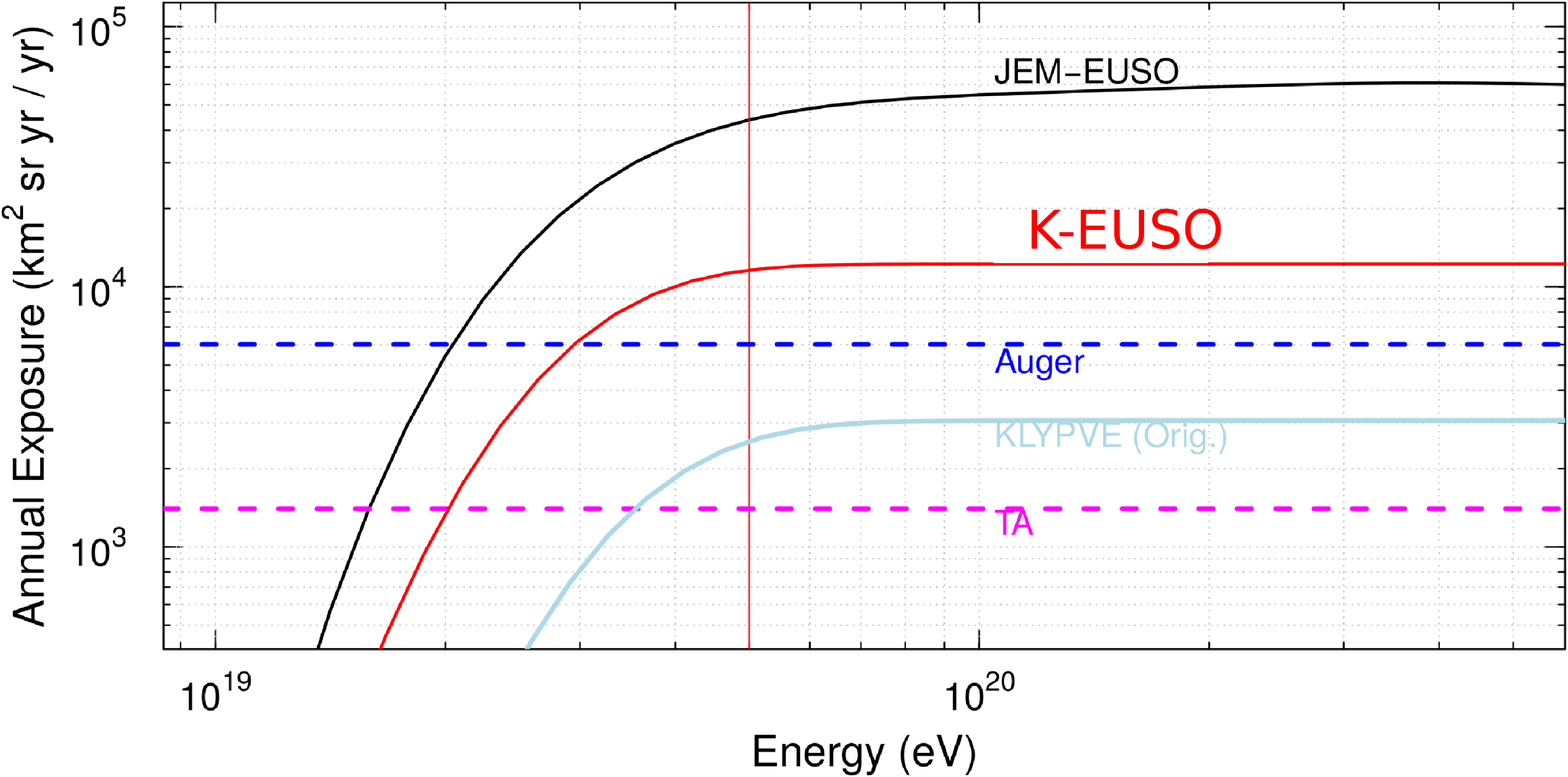}
\caption{the K--EUSO detector exposure curve compared to JEM--EUSO, KLYPVE and ground based detectors \cite{K-EUSO}}
\label{fig11}
\end{figure}
The detector is planned to fly in the early 2020s and to be attached to the Russian section of the ISS. 
The detector will be the first one capable of doing full scale UHECR science from space through fluorescence.
We see in  Fig. \ref{fig11} the comparison between the exposure curves of several existing and planned detectors compared to K--EUSO.
The main goal of this detector is the study of both small and large scale anisotropies, in particular the investigation of the TA hot spot, and of the differences between north and south hemisphere, thanks to its roughly uniform full sky coverage. This represents a major issue and challenge of the current research in the field of UHECRs.

\section{Conclusions}
We gave a brief introduction to the JEM--EUSO program. All the pathfinders were presented and for the already existing ones a brief summary of their main results has been given. 
The collaboration is on its path for the development of space--based UHECR detectors. 
The capability of the JEM--EUSO electronics to trigger CR--like events has been proven together with the technological readiness of the detector. 
The triggered events are used to develop the reconstruction of their direction. 
The response of the JEM--EUSO detector with respect to the background is being studied. The instrument is being calibrated with respect to ground based observatories. 
Within 2017 the collaboration will perform autonomous fluorescence CR detection from balloon and a complete mapping of the UV background from space. 
Such achievements will be useful to build a detector capable of detecting CR from space. K--EUSO should mark the first significant step, providing for the first time a full--sky coverage with a single instrument, with Auger--like statistics in the GZK energy range.

\bigskip
\begin{acknowledgments}
This work has been partially funded by the Italian Ministry of Foreign Affairs and International Cooperation
\end{acknowledgments}

\bigskip 

\end{document}